\documentclass{ws-mpla}
\usepackage[super]{cite}
\usepackage{graphicx}

\newcommand{\li}{${}^{8}{\rm Li}\ $}

\begin{document}

\markboth{H. Kawamura}
{A new measurement of electron transverse polarization in polarized nuclear $\beta$ decay}

\catchline{}{}{}{}{}

\title{
A new measurement of electron transverse polarization in polarized nuclear $\beta$ decay
}

\author{H.~Kawamura$^{1,4}$\thanks{Present address: {\it Frontier Research Institute for Interdisciplinary Sciences, Tohoku University, Sendai, Miyagi 980-8578, Japan}, and {\it Cyclotron and Radioisotope Center, Tohoku University, Sendai, Miyagi 980-8578, Japan }}, 
T.~Akiyama$^1$,
 M.~Hata$^1$, 
Y.~Hirayama$^2$, 
M.~Ikeda$^1$, 
Y.~Ikeda$^1$, 
T.~Ishii$^3$, 
D.~Kameda$^4$, 
S.~Mitsuoka$^3$, 
H.~Miyatake$^2$, 
D.~Nagae$^3$, 
Y.~Nakaya$^1$, 
K.~Ninomiya$^1$, 
M.~Nitta$^1$, 
N.~Ogawa$^1$, 
J.~Onishi$^1$, 
E.~Seitaibashi$^1$, 
S.~Tanaka$^1$, 
R.~Tanuma$^1$, 
Y.~Totsuka$^1$, 
T.~Toyoda$^1$, 
Y.~X.~Watanabe$^2$, 
J.~Murata$^1$}

\address{$^1$Department of Physics, Rikkyo University, Tokyo 171-8501, Japan \\
$^2$Institute of Particle and Nuclear Studies, High Energy Accelerator Research Organization (KEK), Ibaraki 305-0801, Japan \\
$^3$Japan Atomic Energy Agency (JAEA), Ibaraki 319-1195, Japan \\
$^4$Nishina Center for Accelerator-Based Science, RIKEN, Wako, Saitama 351-0198, Japan}

\maketitle

\pub{Received (Day Month Year)}{Revised (Day Month Year)}

\begin{abstract}
The Mott polarimetry for $T$-Violation (MTV) experiment tests time-reversal symmetry in polarized nuclear $\beta$ decay by measuring an electron's transverse polarization as a form of angular asymmetry in Mott scattering using a thin metal foil.
A Mott scattering analyzer system developed using a tracking detector to measure scattering angles offers better event selectivity than conventional counter experiments.
In this paper, we describe a pilot experiment conducted at KEK-TRIAC using a prototype system with a polarized \li beam. 
The experiment confirmed the sound performance of our Mott analyzer system to measure $T$-violating triple correlation ($R$ correlation), and therefore recommends its use in higher-precision experiments at the TRIUMF-ISAC. 
\end{abstract}

\ccode{11.30.Er; 23.40.-s; 24.80.+y; 29.40.Gx}

\section{Introduction}
Since nearly all experimental results in particle physics uphold the Standard Model of particle physics, new phenomena indicating the existence of physics beyond the model remain in demand.
Among such phenomena, a large $CP$- or $T$-violation that could explain considerable matter-antimatter asymmetry observed in our universe might exist.
In response, we tested time-reversal symmetry in nuclear $\beta$ decay in search of a large $T$-violation beyond the Standard Model.

In nuclear $\beta$ decay, a parity-violating phenomenon is known as a correlation between nuclear spin and electron momentum, defined as $A$ correlation in decay rate function, $\omega$; 
\begin{eqnarray}
\label{Eq_betaR}
\omega&\propto &
1 + A \frac{ \vec{p}_e }{ E_e } \cdot \frac{ \langle \vec{J} \rangle }{ J }
+ D \frac{ \langle \vec{J} \rangle  }{ J } \cdot \left( 
\frac{ \vec{p}_e }{ E_e } \times \frac{ \vec{p}_\nu}{ E_\nu }
\right) \nonumber \\
&+&  R \vec{\sigma}_e \cdot \left( 
\frac{ \langle \vec{J} \rangle  }{ J }\times\frac{ \vec{p}_e }{ E_e }
\right)
+ N \vec{\sigma}_e \cdot \frac{ \langle \vec{J} \rangle }{ J } 
\cdots,
\end{eqnarray}
in which, $E_{e(\nu)}$, $\vec{p}_{e(\nu)}$, and $\vec{\sigma}_{e(\nu)}$ are the energy, momentum, and spin of the electron (neutrino), respectively \cite{Jackson1957, Jackson1957NP, Severijns2006}, and $\langle \vec{J} \rangle$ is the parent nuclear spin polarization. 

However, two other correlations also violate time-reversal symmetry.
The first, $D$ correlation is a triple vector correlation of nuclear spin polarization, electron momentum, and neutrino momentum, that does not violate parity symmetry. 
The other correlation, the $R$ correlation, among nuclear spin polarization, electron momentum, and its spin also violates parity symmetry. 
As such, the study of $R$ correlation initially seems to compete with studies of electric dipole moments (EDM), and indeed, search sensitivities of new theoretical models are discussed that compare $R$, $D$ and EDM \cite{Vos2015,Yamanaka2014}. 
For example, in an R-parity violating SUSY model, the $R$ correlation can be large without conflicting with EDMs because the coupling dependence of the correlation differs from that of EDMs\cite{Yamanaka2014}. 
In that sense, $R$ correlation studies may not compete with EDM studies, but complement them instead. 

Regarding the $R$ correlation, a neutron experiment yielding 
$R(n) = (4 \pm 12 \pm 5)\times10^{-3}$
~\cite{Kozela2012}, a \li experiment yielding
$R(^8Li) = (0.9\pm2.2)\times10^{-3}$
~\cite{Huber2003}, a $\mu^+$ experiment yielding
$\langle P_{\rm T_2} \rangle = (-3.7\pm7.7\pm3.4)\times10^{-3}$ 
~\cite{Danneberg2005}, and
other measurements \cite{Cleland1972, Schneider1983} with low precision have all been reported, each with results consistent with time-reversal symmetry. 
We therefore aimed to establish a new technical approach for a new $R(^8Li)$ measurement to be performed at TRIUMF-ISAC, with a sensitivity expected to approach a precision of $10^{-4}$ on $R$.
Two key beam parameters to perform a high precision the $R$ measurement are, 1. a high intensity beam which leads a high statistical precision, and 2. a high beam polarization which enables us to reach high experimental sensitivity.
At TRIUMF, a 70\% polarized $10^7$ pps \li beam is available, which is enabled by laser optical pumping technique \cite{MurataINPC2013, Murata2014}.
At TRIUMF, we expect to perform a test at a precision approximate to the predicted final state interaction level \cite{Huber2003}. 
With that motivation, we performed a pilot experiment at KEK-TRIAC \cite{Watanabe2007}, for which an 8\% polarized $10^5$ pps \li beam was made available for related demonstration measurements~\cite{Kawamura2010}. 
A tilted foil technique was used at TRIAC, which can produce relatively low beam polarization comparing to the optical pumping technique.

We measured the transverse polarization of electrons emitted in polarized \li $\beta$ decay as left-right scattering asymmetry in backward Mott scattering using thin metal foil. 
Mott scattering has an analyzing power on electrons' transverse polarization \cite{Sromicki1999} that has a maximum figure of merit at a backward scattering angle.
Another \li measurement at the Paul Scherrer Institute was performed with the same Mott polarimetry using a plastic scintillation counter array~\cite{Huber2003}.
By extension, we developed a new tracking detector system designed to perform a new Mott polarimetry in order to improve event-detecting reliability and suppress systematic effects in simple counter measurements. 
Inspired by a neutron experiment~\cite{Kozela2012} performed using an electron-tracking detector, as in our approach, for free neutrons, we also measured the Mott scattering of electrons emitted from stopped \li nuclei on a beam stopper by using the tracking detector for the first time. 

\section{Experiment at KEK-TRIAC}
We performed our experiment during 3 days at KEK-TRIAC. 
After the JAEA tandem accelerator irradiated a 68-MeV ${}^{7}{\rm Li}$ beam onto a 99\% enriched ${}^{13}{\rm C}$ graphite disk production target, \li was produced via a ${}^{13}{\rm C} ({}^{7}{\rm Li}, {}^{8}{\rm Li}) {}^{12}{\rm C}$ reaction and separated from the primary beam by JAEA-ISOL~\cite{Watanabe2007}. 
By following a tilted foil method \cite{Hirayama2012}, we then vertically polarized the \li beam, which we stopped at a platinum foil stopper to which magnetic holding fields were applied using permanent magnets. 
Since the beam was diffused via multiple scattering in the tilted foils before entering the stopper, we used an annealed platinum stopper 20~$\mu$m thick and sized 6 $\times$ 7.5~cm to receive the scattered beam.
The typical intensity of the \li beam was $10^5$~pps. 
We measured \li beam polarization as $(8.0\pm1.6)\%$, which yielded an \li sample polarization of $(5.0\pm1.0)\%$ after spin relaxation in the stopper material \cite{Hirayama2012}. 

To cancel the detector's efficiency difference for left- and rightward Mott scattering, we flipped the direction of the nuclear polarization every 5 min. 
We performed beam spin flipping by rotating the tilting foils in the reverse direction, which took 20~s.
We next set the measuring time window to be 300~s. 
Table~\ref{Tab_8Lipara} presents the beam parameters.
 
\begin{table}[!htpb]
\begin{center}
\tbl{\li beam parameters.}
{\begin{tabular}{rl}
\hline
\hline
  \multicolumn{2}{c}{ ${}^8{\rm Li} \to {}^8{\rm Be} + e^- + \bar{\nu}$} \\
\hline
  Life time                         & $\tau=1.21$~s \\
\hline
  Spin-parity                       & $I^{\pi}=2^+ \to 2^+ $ \\
                                    & pure Gamow-Teller \\
\hline
  Maximum $\beta$ energy            & $E_{\rm max}=12.9645$ MeV \\
\hline
  $\beta$-decay asymmetry parameter & $A=-1/3$ \\
\hline
\end{tabular}}
\label{Tab_8Lipara}
\begin{tabular}{rl}
\hline
  Tilted foil:   & 20-polystyrene films, \\
                 & $t$-30 nm ($\sim3$ $\mu{\rm g}/{\rm cm}^2$) \\
  Tilting angle: & $70^{\circ}$ \\
  Gap of foils:  & 3 mm \\
  Beam purity:   & $\geq99.99\%$ \\
  Beam energy:   & 0.178 MeV/nucleon \\
                 & (duty cycle 50\%) \\
  Typical beam intensity: & $10^5$ pps \\
  Beam polarization: & $(8.0\pm1.6)\%$ \\
  Beam stopper:  & Pt, $t$-20 $\mu$m, annealed \\
  \li sample polarization: & $(5.0\pm1.0)\%$ \\
  Spin relaxation time:  & 2 s \\
  Magnetic field:& \hspace{0.3em}\raisebox{0.4ex}{$>$}\hspace{-0.75em}\raisebox{-.7ex}{$\sim$}\hspace{0.3em}500 Gauss \\

\hline
\hline
\end{tabular}
\end{center}
\end{table}

\begin{figure}[htb]
  \begin{center}
  \includegraphics[clip, width=5.0cm]{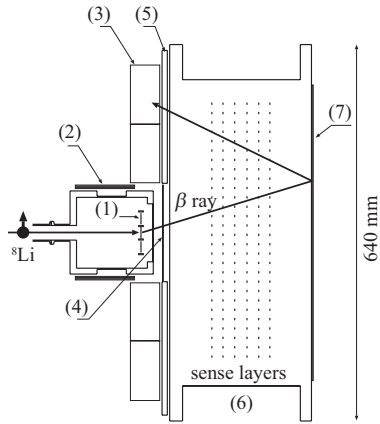}
  \caption{Experimental setup of the Mott analyzer. 
(1) Platinum beam stopper, 
(2) $\beta$ asymmetry counter, 
(3) $E$ counter, 
(4) $\delta$ counter, 
(5) $\Delta$ counter, 
(6) multiwire drift chamber, 
(7) Mott analyzer foil. 
$\beta$ asymmetry counters measure the nuclear polarization. 
The first-level trigger is the coincidence of $\delta$ and $\Delta$ signals, whereas the second=level 2 trigger is generated using MWDC hit number information in the FPGA.  
$E$ counters measure electron energy. 
}
  \label{figure1}
  \end{center}
\end{figure}

\begin{figure}[htb]
  \begin{center}
  \includegraphics[clip, width=8.5cm]{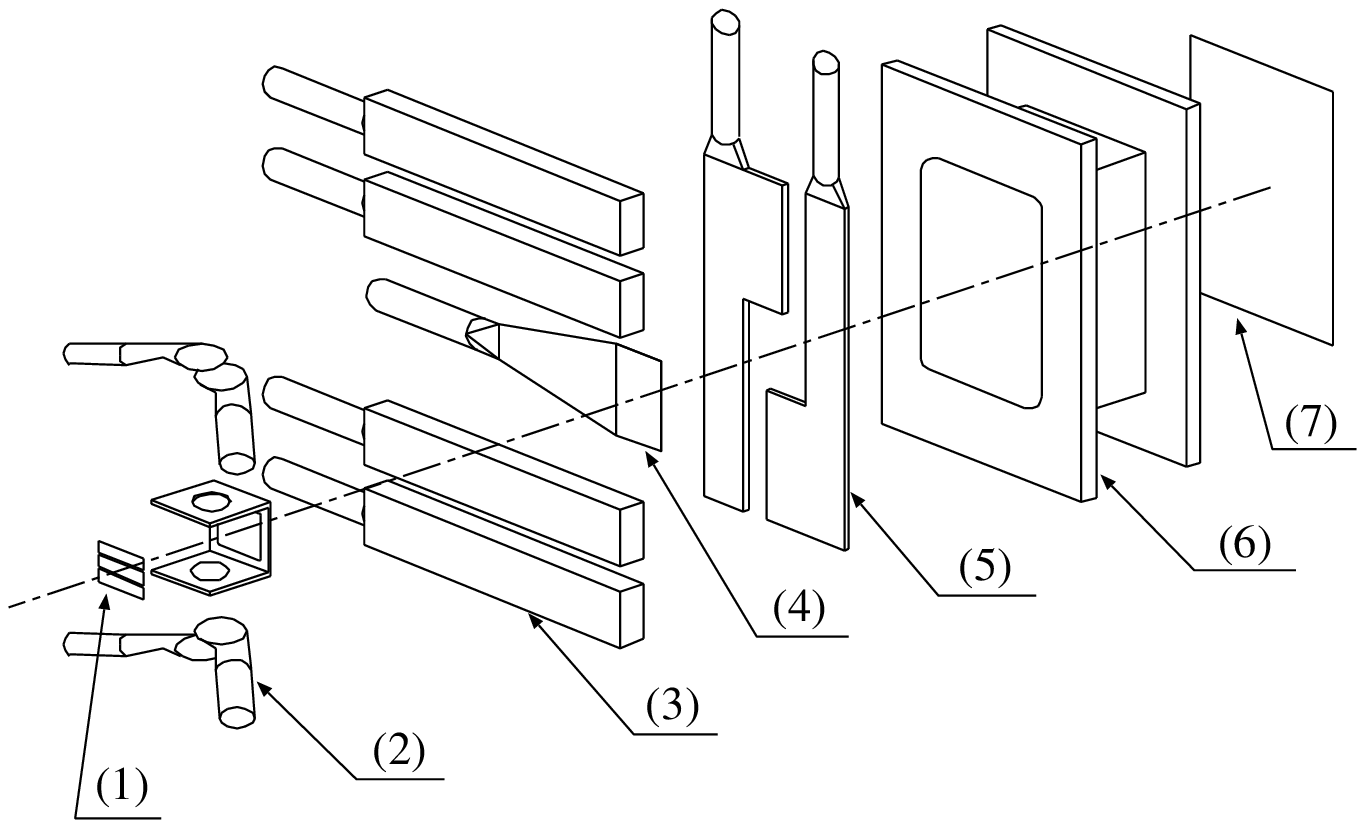}
  \caption{Expanded view of the setup; numbers are the same as those in Figure~\ref{figure1}.}
  \label{figure2}
  \end{center}
\end{figure}

Electrons emitted from \li on the platinum stopper placed in a vacuum chamber passed through an aluminum vacuum window 300~$\mu$m thick and entered the tracking detector, as shown in Figure \ref{figure1}. 
We used a planar multiwire drift chamber (MWDC) as a three-dimensional tracking detector, which operated with P10 gas (Ar 90\%+${\rm CH}_4$ 10\%) under atmospheric pressure. 
We placed an analyzer foil on the rear side of the MWDC, as shown in Figure \ref{figure2}. 
We used a lead foil 100~$\mu$m thick as the analyzer foil, which we supported with a thin plastic film. 
With that configuration, both incident tracks going to the analyzer foil and backward scattered tracks from the foil were detectable. 
We determined the scattering angles of these so-called V tracks event by event. 
 

\begin{figure}[htb]
  \begin{center}
  \includegraphics[clip, width=8.0cm]{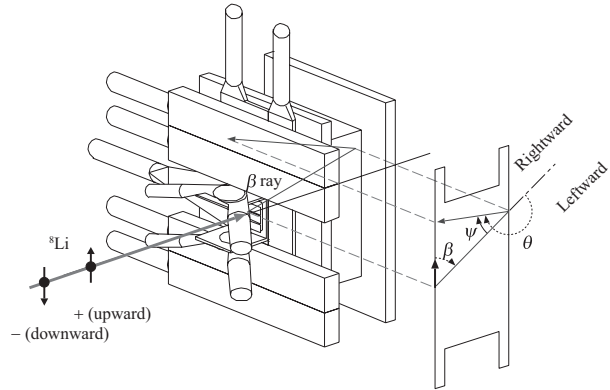}
  \caption{Reconstructed tracks in three dimensions projected onto the plane, which defines the decay angle $\beta$, scattering angle $\theta$, and opening angle $\psi$. 
Beam spin ``$+$'' or ``$-$'' is defined as the upward or downward direction, respectively. }
  \label{figure3}
  \end{center}
\end{figure}

We measured electron hit positions in the six sense layers (XX'UU'VV') of the MWDC, which we used to reconstruct the three-dimensional tracks. 
We set the anode direction on the UU' and VV' layers to stereo angles of $\pm 15.9 ^{\circ}$ to the XX' layer; 
the anode cell size was 2 $\times$ 2 cm for all sense layers. 
We next projected the reconstructed three-dimensional V tracks onto a plane parallel to nuclear polarization direction, as shown in Figure~\ref{figure3}.
We defined the opening angle of the projected V track as the projected Mott scattering angle, $\psi$, in a subsequent analysis, as described in the following. 
Therein, $\psi=\theta-180^{\circ}$, as determined by using the projected scattering angle $\theta$, as defined in Figure \ref{figure3}; 
$-180^{\circ}\leq\psi<180^{\circ}$ and 
$0^{\circ}\leq\theta<360^{\circ}$. 
In that projected plane, we defined the electron's emission angle from the nuclear polarization direction as $\beta$. 
To select rare backward scattering events from the dominant straight background, we built a field-programmable gate array (FPGA)-based triggering system to select events containing at least two independent hits in each sense layer. 
The FPGA-based second-level trigger checked MWDC's anode hit number after obtaining a first-level trigger consisting of a plastic scintillation counter logic requesting hits on $\delta$ and $\Delta$ counters. 
Afterward, we rejected background events that did not have a scattering vertex position at the analyzer foil. 

\begin{figure}[htb]
  \begin{center}
  \includegraphics[clip, width=6.0cm]{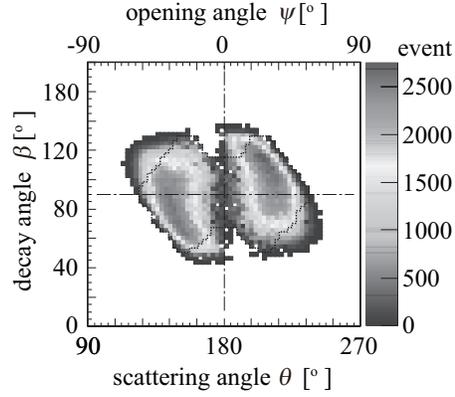}
  \caption{Two-dimensional distribution of scattering angle $\theta$ and decay angle $\beta$; 
events inside the dotted line are used for asymmetry analysis. 
This cut is to keep left-right symmetric acceptance. The cutting boundary was set to maximize the acceptance, to keep both $(\theta<180^{\circ})$ and $(\theta>180^{\circ})$ regions have nonzero detected events, at a same $\beta$ region.
}
  \label{figure4}
  \end{center}
\end{figure}

\begin{figure}[htb]
  \begin{center}
  \includegraphics[clip, width=6.0cm]{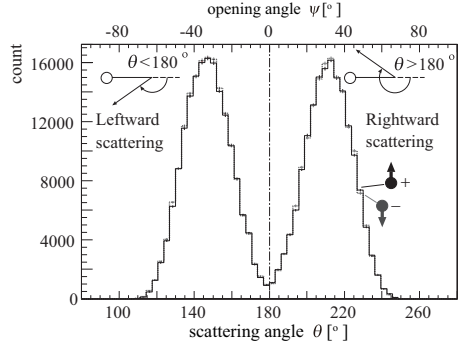}
  \caption{Angular distribution of scattering angle, $\theta$; 
the solid line indicates the spin $+$ beam, and the dotted line indicates the $-$ case. }
  \label{figure5}
  \end{center}
\end{figure}

Figure~\ref{figure4} shows the two-dimensional distribution of the Mott scattering angle $\psi$ and the decay angle $\beta$. 
The characteristic shape of this distribution shows an acceptance of the detector. 
The Mott scattering angular distribution is plotted in Figure~\ref{figure5} for beam spin ``$+$'' and ``$-$'' cases upward and downward along the vertical axis, respectively. 
Without using an analyzer foil, we estimated background events not scattered by the analyzer foil to be $\sim 6\%$ of the total number of events. 
We subtracted that off-foil background contribution in offline analysis. 
In total, we obtained $6.7\times10^5$ V track events, which we selected from the $1.1\times10^7$ recorded events and used in subsequent analysis.


\section{Data Analysis and Results}
We obtained the two-dimensional yield distribution of $\beta$ versus $\psi$, as shown in Figure \ref{figure4}, as
 $N^+(\psi,\beta)$ for the beam spin $+$ and as $N^-(\psi,\beta)$ for the beam spin $-$. 
We then obtained the left- and rightward Mott scattering ratio, $r$, as 
\begin{center}
\begin{eqnarray}
r^+(|\psi|,\beta)&=&N^+(-|\psi|,\beta)/N^+(+|\psi|,\beta),\\
r^-(|\psi|,\beta)&=&N^-(-|\psi|,\beta)/N^-(+|\psi|,\beta),
\end{eqnarray}
\end{center}
for the spin $+$ and $-$ cases. 
Therein, $+|\psi|$ and $-|\psi|$ imply right- and leftward scattering events, respectively, at opening angle $|\psi|$. 
To cancel the detector's left or right efficiency difference, we defined double ratio $d$ as
\begin{equation}
d(|\psi|,\beta)=\frac{r^+(|\psi|,\beta)}{r^-(|\psi|,\beta)}.
\end{equation}
We next defined Mott asymmetry $A_{sym}^{Mott}$ as
\begin{equation}
A_{sym}^{Mott}(|\psi|,\beta)=\frac{\sqrt{d(|\psi|,\beta)}-1}{\sqrt{d(|\psi|,\beta)}+1}.
\end{equation}
Ideally, the expected counting yield is expressible by supposing the $A$ and $R$ correlations:
\begin{eqnarray*}
N^{+}_{-} (+|\psi|, \beta) \propto  1 \pm P_n \beta_e A \cos\beta \mp P_n \beta_e S(|\psi|,\beta_e) R \sin\beta,  \\
N^{+}_{-} (-|\psi|, \beta) \propto  1 \pm P_n \beta_e A \cos\beta \pm P_n \beta_e S(|\psi|,\beta_e) R \sin\beta, 
\end{eqnarray*}
in which $\beta_e$ is $v_e/c$ with electron velocity $v_e$, $S(|\psi|,\beta_e)$ is the analyzing power of Mott scattering, and $P_n$ is the nuclear polarization of ${}^{8}\textrm{Li}$.
More realistically, it appears as
\begin{eqnarray*}
N^{+}_{-} (+|\psi|, \beta) \propto  1 \pm A f(\beta_e,\beta) \mp R S(|\psi|,\beta_e) g(\beta_e,\beta),  \\
N^{+}_{-} (-|\psi|, \beta) \propto  1 \pm A f(\beta_e,\beta) \pm R S(|\psi|,\beta_e) g(\beta_e,\beta),
\end{eqnarray*}
using the distribution functions $f(\beta_e,\beta)$ and $g(\beta_e,\beta)$, which include a realistic efficiency distribution $\epsilon(\psi,\beta)$ including the nonuniformity of the analyzer foil.
In that case, we expected the double ratio to be 
\begin{equation}
d\equiv\frac{1+Af+RSg}{1+Af-RSg}\cdot \frac{1-Af+RSg}{1-Af-RSg}
\sim1+\frac{4RSg}{1-(Af)^2} \;\;\;\; (R\ll1)
.
\label{def-d}
\end{equation}
In our case, in the condition of $A=-1/3$ and $P_n=5\%$, the correction factor is negligible as  $1/(1-(Af)^2) -1 <10^{-3}$.
\begin{equation}
A_{sym}^{Mott}\sim\frac{d-1}{4}\sim\frac{RSg}{1-(Af)^2}  \sim RSg \;\;\;\; (R\ll1)
.
\label{asym-approximation}
\end{equation}

Therefore, we could regard Mott asymmetry $A_{sym}^{Mott}$ as nearly proportional to the $R$ coefficient within its poor relative precision.

\begin{figure}[htb]
  \begin{center}
  \includegraphics[clip, width=6.0cm]{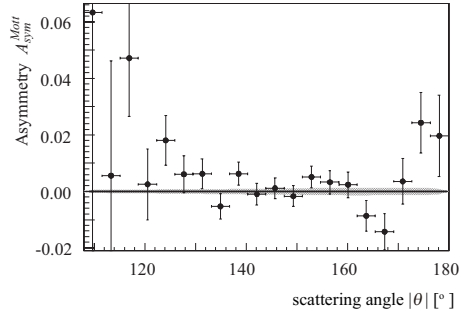}
  \caption{The scattering angle dependence of Mott asymmetry; 
a horizontal line of nearly $A_{sym}^{Mott}\sim0$ shows the best fit obtained, with the error indicated in the shaded area. 
Note that the fitting result was obtained from a direct two dimensional analysis in $\psi$ and $\beta$, not from the projected data shown in this figure.}
  \label{figure6}
  \end{center}
\end{figure}

In our analysis, we obtained the $R$ coefficient by fitting the obtained  $A_{sym}^{Mott}$ as a two-dimensional function of $\beta$, $\psi$; 
however, $\beta$ dependence was small because $Af$ in Equation (\ref{def-d}) appeared only in the second-order contribution to the solution of $RSg$, which was strongly suppressed. 
Therefore, an analysis that considers only $\psi$ dependence may be a good approximation; 
however, we performed direct analysis in our work. 
The angular distribution of Mott scattering appears in Figure \ref{figure6}, following the integration of the decay angle $\beta$. 
As for $\beta_e$ dependence, instead of directly treating the $\beta_e$ value for each event, we used the mean value of $\beta_e$. 

To obtain the $R$ coefficient, we required analyzing power $S(|\psi|,\beta_e)$, which Sherman has calculated in detail \cite{Sherman1956}. 
We used an effective analyzing power, $S_{eff}(|\psi|)$, after integrating the electron energy dependence, thereby correcting for the foil thickness effect due to depolarization inside the lead foil \cite{Sromicki1999} and correcting the tilting effect of the Mott scattering plane.
The $\theta$-dependent shape of $S_{eff}(|\psi|)$ was also corrected in consideration of the angular resolution of the MWDC, which was approximately $10^{\circ}$. 
We moreover corrected the depolarization effect of the electrons \cite{Koks1976} that passed through the stopper and detector materials. 

\begin{figure}[htb]
  \begin{center}
  \includegraphics[clip, width=6.0cm]{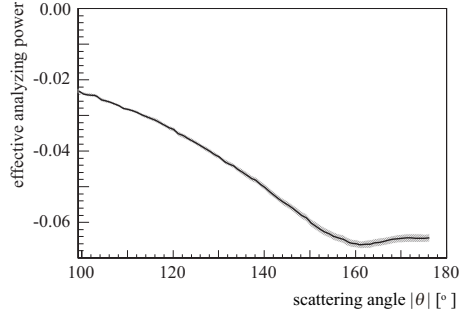}
  \caption{ Scattering angle dependence of the effective analyzing power; 
shading indicates systematic uncertainty. }
  \label{figure7}
  \end{center}
\end{figure}

Figure~\ref{figure7} shows the effective analyzing power $S_{eff}$ after correcting all effects to yield a case of applied $\beta$ integration. 
We integrated the $\beta_e$ and $\beta$ dependences to identify that effective analyzing power; 
$\psi$ integration defined an average value of the effective analyzing power as 
$
\langle S \rangle = -0.065\pm0.020.
$

By using the two-dimensional effective analyzing power $S_{eff}(|\psi|,\beta)$ with the experimental results of $A_{sym}^{Mott}(|\psi|,\beta)$, we obtained $R=0.02\pm0.40$ as the best fit parameter. 
The result contained errors from not only $S_{eff}$, but also the nuclear polarization $P_n$. 
We also considered several systematic errors, as follows. 

{\it Electron energy threshold. }
Analyzing power $S$ depends upon the scattering angle and electron energy. 
For simplicity's sake, we did not use the measured energy value in analysis. 
The energy threshold of the counters corresponded to the minimum electron energy of $(2\pm1)$~MeV before losing its energy inside the materials. 
Such uncertainty of the energy threshold induced an error in analyzing power $S_{eff}$,  which prompted a systematic error of $1.4\times 10^{-3}$ on $R$. 

{\it Beam polarization tilting. }
We assumed the analyzing plane projected onto the V tracks to be parallel to nuclear polarization. 
However, if we tilted the nuclear polarization from the vertical direction around the beam axis, then we had to consider a sizable contribution from the $N$ correlation defined in Equation (\ref{Eq_betaR}). 
If we defined the tilting angle $\alpha$ as the beam polarization direction around the beam axis from the upward direction, then we expected 
$$
A_{sym}^{Mott} \sim P_n \beta_e S(\theta) \sin(\beta) \left[ 
R \cos(\alpha) + N \sin(\alpha)
\right]
$$
in which $\alpha=0$ represents the ideal vertical ``$+$'' configuration. 
In the case that $\alpha\neq 0$, then the contribution from $N$ needs correction. 
We estimated the $N$ parameter in the standard model as $N=-\sqrt{ 1-\alpha_F^2Z^2 }m_e/E_e\cdot A=0.04$ when we assumed an average electron energy of 4~MeV from \li decay. 
We estimated tilting angle $\alpha$ with an alignment error of $\pm1.2^{\circ}$, in which the mounting precision of the tilted foil of 
$\pm1^{\circ}$ represented the dominant uncertainty. 
The estimated systematic error on $R$ due to $N$ correlation was $8.3\times10^{-4}$. 

{\it $\beta$-decay asymmetry. }
If efficiency was not perfectly stable, but correlated with radiation intensity, then a fake asymmetry was observable without real $R$ correlation due to the anisotropic change in electron flux, which was due to the parity violating $\beta$ asymmetry.
Moreover, accidental coincidence between two different event tracks that were wrongly recognized as V track events may generate additional fake asymmetry. 
Such systematic effects cannot be negated by beam spin flipping. 
We studied this fake asymmetry effect in another experiment at TRIUMF-ISAC as MTV Run-II~\cite{Murata2015}. 
By extrapolating from the nuclear polarization of 58\% at TRIUMF and 5\% at TRIAC, we can estimate the expected contribution of the effect to the $R$ coefficient to be 0.14 in the present TRIAC experiment. 
However, given the difficulty of estimating the ambiguity of the correction without performing a dedicated measurement, we added the corrected amount to the error, which thus became overestimated. 

Ultimately, the resulting $R$-coefficient was
\begin{equation}
R = 0.02 \pm 0.40_{\rm stat} \pm 0.15_{\rm syst}.
\end{equation}
Key parameters to extract this result are summarized in Table~\ref{Tab_result}. 
Error budgets appear in Table~\ref{Tab_RR}.

\begin{table}[h]
\tbl{Summary of the present experiment. Two dimensional analysis on $(\psi,\beta)$ is performed to extract $R$.}{
\begin{tabular}{l|l|l}
\hline
\multicolumn{2}{l|}{$R$ correlation}    & $R = 0.02 \pm 0.40_{\rm stat} \pm 0.15_{\rm syst}$ \\
\hline
\hline
\multicolumn{2}{l|}{observed Mott asymmetry} &  $A_{sym}^{Mott}(|\psi|,\beta)$ (1d example : Figure \ref{figure6}) \\
\hline
\multicolumn{2}{l|}{measured sample polarization} &  $P_n = (5.0\pm1.0)\%$\\
\hline
\multicolumn{2}{l|}{calculated effective analyzing power} &  $S_{eff}(|\psi|,\beta)$ (1d example : Figure \ref{figure7}) \\
\hline
\end{tabular}
}\label{Tab_result}
\end{table}

\begin{table}[h]
\tbl{Error budgets of the obtained $R$-parameter.}{
\begin{tabular}{l|l|l}
\hline
\multicolumn{2}{l|}{statistical error}    & 0.40 \\
\hline
\hline
\multicolumn{2}{l|}{systematic error} & 0.15 \\
\hline
           & \li polarization       & 0.004 \\
           & effective analyzing power   & 0.006 \\
            & energy threshold            & 0.0014 \\
            & beam polarization tilting   & 0.0008 \\
            & $\beta$-decay asymmetry     & 0.14 \\
\hline
\end{tabular}
}\label{Tab_RR}
\end{table}

\section{Conclusion}
The MTV project represents an ongoing search for nonzero $R$ correlation in the $\beta$ decay of polarized unstable nuclei at a precision reaching the level of the final state interaction. 
We performed the first experiment at KEK-TRIAC using the new event-by-event tracking Mott polarimetry technique, and we found the $R$ coefficient to be consistent with zero when the statistical error was dominant. 
We thus successfully performed the new $T$-violation experiment using the new tracking principle. 


\section*{Acknowledgments}
The authors thank the technical support at JAEA accelerator group and the staff at TRIAC. This work was supported by JSPS KAKENHI Grant-in-Aid for Young Scientists (A) 18684010 and 21684012, Scientific Research (B) 25287061, and Rikkyo SFR (Rikkyo University Special Fund for Research).

\end{document}